# A Group Key Management Protocol Based on Weight-Balanced 2-3 Tree for Wireless Sensor Networks

Lin Yao*, Bing Liu*, Feng Xia*, Guo-Wei Wu* and Qiang Lin**

*School of Software, Dalian University of Technology, Dalian 116620, China*
*E-mail: f.xia@ieee.org*

**Information Engineering Institute, Dalian Jiaotong University, Dalian 116028, China*

## Abstract

Multicast in Wireless Sensor Networks (WSNs) is an attractive mechanism for delivering data to multiple receivers as it saves bandwidth. To guarantee the security of multicast, the group key is used to encrypt and decrypt the packages. However, providing key management services in WSNs is complicated because sensor nodes possess limited resources of computing, storage and communication. To address the balance between security and limited resources, a multicast group key management protocol based on the weight-balanced 2-3 tree is proposed to generate, distribute, and update the group key securely and efficiently. The decentralized group key management method is employed. A weight-balanced 2-3 key tree is formed in every subgroup. Instead of using the conventional symmetric and non-symmetric encryption algorithms, the Maximum Distance Separable (MDS) code technique is used to distribute the multicast key dynamically. During the key updating, a series of adjustment rules are summarized to keep the tree weight-balanced, where pseudo-nodes as leaves are added to reduce the computation and communication complexity. Compared with some other group key management protocols, our scheme shows higher superiority on security and performance.

**Key Words**: Wireless sensor networks, Group key management, Multicast, 2-3 tree

## 1. Introduction

In a WSN, multicast is a more efficient method of supporting group communication than unicast, which can save bandwidth and energy largely. Because sensor nodes suffer from limited resources such as low computation capability, small memory, limited energy resources and low bandwidth, multicast is widely used in sensor-based applications such as environment monitoring, health-care monitoring, etc [1]. In general, multicast is far more vulnerable than unicast because the transmission takes place over multiple network channels [6]. It is an important issue to assure the security for the multicast communication in WSNs. Due to the resource constraints of sensor nodes, security in WSNs imposes several challenges which are more complex than those in the other traditional networks [2]. At the same time, the large number of sensor nodes employed to monitor critical parameters, the lack of a specific network architecture or infrastructure, and the frequent topology changes due to the nodes mobility also bring the challenges when we design the group key management protocol [3].

To overcome these issues, the first security challenge is to provide an effective method to protect information through encryption, i.e., every message must be encrypted with the Group Key (GK). GK is known only by legal group members. The second security challenge is that GK should be changed when the nodes join or leave the group. Changing GK can not only prevent a new member from deciphering previous messages encrypted with the old GK, but also prevent a leaving or expelled member from accessing the subsequent messages. The third security challenge is to consider the balance between the security and the limited resources of sensors.

To sum up the above arguments, the multicast key agreement protocol in WSNs should meet the following requirements:

(1) *Backward secrecy* [4, 5]: It is used to prevent a new member from decoding messages exchanged before it joins the group.

(2) *Forward secrecy* [5]: It is used to prevent a leaving or expelled group member from continuously accessing the group's communication.

(3) *Key Independence*: GK should be random enough and there is no any correlation between each other.

(4) *Low computation, storage and communication cost*: As sensors possess limited resources, the resources should be used as few as possible when a GK is computed.

In this paper, we propose a tree-based hierarchical group key management scheme for WSNs to meet the above requirements. The decentralized group key management method is employed and the main contributions are as follows.

(1) *The key tree is used to efficiently compute and update GK*. In our paper, the network model is shown in Fig. 1. The Base Station (BS) acts as the whole tree root in the layer one and is responsible for managing the whole tree. In the layer two, subgroups are divided according to the domain in which every Sink Node (SN) locates. SN stands for the sub-tree root and is responsible for managing its sub-tree. The sensor nodes except SNs act as the leaves in layer two. The other nodes except BS, SNs and leaves are logic nodes in order to form a weight-balanced 2-3 tree, where each internal node has degree 2 or 3. A weight-balanced 2-3 tree can achieve the lowest worst-case cost when membership changes [28]. According to the network model, the decentralized group key management method is employed in our paper. The keys of all nodes form a key tree. GK serves as the root and the subgroup keys correspond to the SN nodes. Each member stores all the keys along the path from itself to the tree root.

(2) *The Maximum Distance Separable (MDS) code technique is used to distribute GK dynamically*. Compared with the symmetric or asymmetric cryptography mechanisms, MDS codes [24] can reduce the computation load and the energy consumption of each group member, which also maintains low and balanced communication complexity and storage complexity for secure multicast key distribution.

(3) *A series of adjustment rules for the 2-3 trees are set to keep the tree balanced when the membership changes*. In order to reduce the computation and communication complexity, pseudo-nodes as leaves are added to keep the tree balanced if necessary.

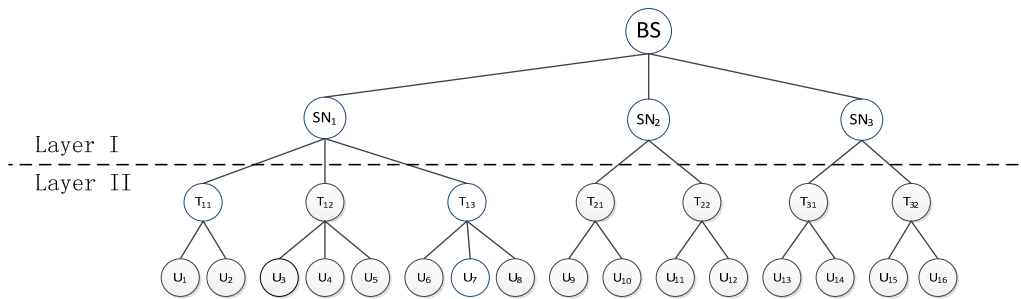

Fig. 1. The network model

The remainder of this paper is organized as follows. In Section 2, we discuss related work. Section 3 describes the proposed group key management protocol based on the weight-balanced 2-3 tree in detail. The security and performance analysis are given in Section 4. Section 5 concludes the paper.

## 2. Related Work

Group key management creates, distributes and updates GK for the group members. The group key management protocols can be divided into three main categories [8]: the centralized group key management protocol, the distributed group key management protocol and the decentralized group key management protocol.

In the centralized group key management protocols [8, 9, 17, 18, 19], a BS is employed to control the whole group as a group controller (GC). BS assigns a secret key to each group member and encrypts GK separately by each member's secret key when a new GK needs regenerating and distributing. Unfortunately, the centralized group key management protocol does not suit the highly dynamic WSNs because frequent joining or leaving of members brings great computation burden on GC and causes the single point of failure easily.

In order to overcome the shortcomings of the centralized group key management protocol, the distributed key management protocol is proposed [10,11,12]. GC does not exist and group members are peer to peer, therefore GK is generated by all the members. A group rekeying scheme (PCGR) for filtering false data was proposed [12], whose basic idea is that the GK can be preloaded to the sensor nodes before deployment and neighbors can collaborate to protect and appropriately use the preloaded keys. However, the authors do not address the group rekeying problem when the group includes sensor nodes separated by multiple hops. A group key distribution via local collaboration protocol (GKD) was proposed [10]. GKD is not only based on pre-distributed personal secrets and broadcast information, but also requires local collaboration among sensor nodes. The network lifetime is divided into time intervals known as sessions. At each session, BS broadcasts to the whole group in order to initiate the group key updating. To derive the GK actually hidden in the broadcast message, a node has to seek trust and exchange secrets with others. Such distributed group key management protocol can avoid the single point of failure, but brings more difficulty on group management owning to lack of the centralized control. As the size of the network increases, the communication and computation costs on each sensor node will also increase [19].

The decentralized group key management protocol as the third approach is proposed [7, 13, 14, 15, 16, 20, 21, 22, 23]. The large group is divided into small subgroups. Each subgroup is independent of each other, and the membership change in a subgroup can be confined locally. Because a WSN usually is divided into several clusters, the protocols for a WSN can be categorized into this approach [7]. The centralized group key management or the distributed group key management protocol can be used in every subgroup. A scalable and efficient master key management protocol was proposed [7]. Two types of keys, the cluster key and the master key, are generated in this protocol. Sensor nodes in the same cluster shares a cluster key with their gateway node and utilize this cluster key to encrypt cluster traffic. Besides, all sensor nodes share a master key with the base station. When group member's move, the communication and computation cost of each component for master key renewing can be confined to the affected cluster. The base station can utilize this master key to securely broadcast the information to all sensor nodes. A novel key agreement protocol based on layer-cluster group model was proposed [20]. In this protocol, the cluster keys are generated based on security elliptic curves, which can offer smaller system parameters, lower power consumptions and faster implementations. A Refined Key Link Tree (RKLT) scheme was proposed [21]. The main contribution of this paper is the introduction of a dirty key path into the key link tree, which starts with a leaf node and ends with the root node if the keys of the auxiliary nodes in this path don't get updated after the leaf node leaves. During the key update, RKLT scheme delays the modification of auxiliary nodes in order to reduce the number of reduplicate key update messages for the same auxiliary nodes, which also brings down the energy cost. A new energy-efficient key management scheme for networks based on secret sharing (GKSS) was proposed [22]. The cluster heads choose the one-order polynomial to calculate the secret shares which are encrypted by the symmetric keys preloaded into the nodes.

The decentralized group key management protocol is a hybrid group key management scheme involving both centralized and distributed key management schemes, which is a kind of more practical key management scheme. But all the protocols mentioned above show defects in computation and storage cost. During the group key distribution scheme, a basic operation is to distribute a piece of secret data to a small group of members, where each shares a key with the subgroup controller. In the existing schemes, this is done by symmetric encryption algorithms. Frequent key update caused by the frequent membership change may bring expensive computation and high storage requirements, which cannot be afforded by sensor nodes with limited resources. It is a critical issue to design a provably secure group key management protocol with practicability, simplicity, and strong security in WSNs.

## 3. Our Group Key Management Protocol

As noted in section 1, our group key management protocol for WSNs is based on weight-balanced 2-3 trees, where the distributed group key management method is employed. BS is responsible for the whole tree. Instead of using the conventional symmetric and non-symmetric encryption algorithms, MDS is employed to distribute the multicast key dynamically. In this section we begin with a brief discussion of the group initialization. Next, we introduce our protocol when membership changes such as join, leave, mergence, and partitions. The authentication between nodes is beyond the scope of this paper. Our main aim is focus on how to manage the GK. For ease of reference, the notations used in the protocol description are listed in Table 1.

### 3.1 Group Initialization

In our protocol, keys are arranged in a logic tree hierarchy. One-way functions are used to compute a tree of keys [21]. Fig. 1 shows a key tree including sixteen members. $K_{BS}$ as GK is shared by all group members. The keys in this tree are computed from the leaves to the root. Each internal node key can be used as a subgroup key for all descendent members of the internal node. $K_{SN_1}$, $K_{SN_2}$ and $K_{SN_3}$ are three subgroup keys. For example, $K_{SN_1}$ is shared by members $u_1$ to $u_8$.

Table 1. Notations of protocol description

| Symbol | Description |
| --- | --- |
| $E_k(m)$ | Encrypt message $m$ using key $k$ |
| GK | Group key |
| T | Internal node in the tree |
| $K_x$ | A key of the node $x$ |
| H | A cryptographically strong hash function |

In all current existing schemes for WSNs, the update of GK is fulfilled by multiple encryption and decryption operations. In order to reduce these operations, MDS codes are used to achieve the dynamic key distribution. MDS codes are a class of error control codes that meet the Singleton bound [24, 25]. The well known Reed-Solomon codes are widely used as a class of MDS codes. Let $GF(q)$ be a finite filed with $q$ elements, an $(n, k)$ linear block code is an error control code which maps from $GF(q)^k$ to $GF(q)^n$. Let $E(\ )$ be an encoding function, $E(m)=c$ is set where $m = m_1 m_2 m_3 ... m_k$ is the original message code and $c = c_1 c_2 c_3 ... c_n$ is its code word block with $k \leq n$. Let $D(\ )$ be a decoding function, $D(c_{i_1} c_{i_2} ... c_{i_k}, i_1, i_2, ... i_k) = m$ is set with $1 \leq i_j \leq n$ and $1 \leq j \leq k$ where $m = m_1 m_2 m_3 ... m_k$ is the original message code. For an $(n, k)$ MDS code, the $k$ original message symbols can be recovered from any $k$ of its code word block.

During the group initialization, every SN assigns its member a different random number $j_i$ and a random hash value $s_i$ such as the hash value of MD5 algorithm securely. The pair $(j_i, s_i)$ is used as $u_i$'s seed key and kept in the local database of SN. BS also assigns a seed key for every SN securely. As shown in Fig. 1, $SN_1$ assigns seed keys for $u_1$ to $u_8$. $SN_1$ maintains a tree whose depth is three. Group members from $u_1$ to $u_8$ are leaf nodes and $T_1$, $T_2$ and $T_3$ are logic nodes. $SN_1$ needs to assign a different position number to every node which can be seen as $j_i$. The root of $u_1$ and $u_2$ is the logic node $T_{11}$. $SN_1$ executes the following steps[25] to compute $K_{T_{11}}$ in Fig. 2(a):

(1) $SN_1$ randomly chooses a fresh element $r$ in $F$, which has not been used to generate previous keys.

(2) $SN_1$ computes $GF(q): c_{j_i} = H(s_i \| r)$ for every leaf $u_i$ with $i = 1...n$.

(3) Using all the $c_{j_i}$ in the step (2), $SN_1$ constructs a code word $c$ of the $(L, n)$ MDS code $C$ and the ($j_i$)th symbol of $c$ is $c_{j_i}$. Since $c$ is an $(L, n)$ MDS code, the code word $c$ is uniquely determined by its $n$ symbols. Using an efficient erasure decoding algorithm for $c$, $SN_1$ can easily calculate the $n$ corresponding message symbols $m_1 m_2 m_3 ... m_n$.

(4) $SN_1$ sets the new session key $K_{T_{11}}$ to be the first message symbol $m_1$.

(5) $SN_1$ multicasts $r$ and $m_2 m_3 ... m_n$ to $u_1$, $u_2$ and $u_3$.

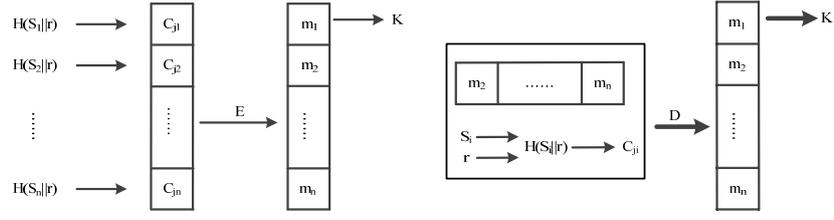

(a) The process of generating the key (b) The process of releasing the key

Fig. 2. The process of generating and releasing the key

Upon receiving $r$ and $m_2 m_3 ... m_n$ from $SN_1$, an authorized group member[30] executes the following steps to obtain the new session key in Fig. 3(b):

(1) Calculate $c_{j_i} = H(s_i \| r)$ with its seed key $(j_i, s_i)$.

(2) Calculate $m_1$ with $c_{j_i}$ and $m_2 m_3 ... m_n$ by the following equation [26].

$$\begin{bmatrix} 1 & j_1 & \cdots & (j_1)^{n-1} \\ 1 & j_2 & \cdots & (j_2)^{n-1} \\ \cdots & \cdots & \cdots & \cdots \\ 1 & j_n & \cdots & (j_n)^{n-1} \end{bmatrix} \begin{bmatrix} m_1 \\ m_2 \\ \cdots \\ m_n \end{bmatrix} = \begin{bmatrix} c_{j1} \\ c_{j2} \\ \cdots \\ c_{jn} \end{bmatrix} \quad (1)$$

(3) Recover $K_T$, i.e., $K_{T_{11}} = m_1$.

$SN_1$ performs the same process to compute $K_{T_{12}}$ and $K_{T_{13}}$ as shown in Fig. 2(a). In order to save the storage, $SN_1$ does not assign random hash values $s_{T_{11}}$, $s_{T_{12}}$, and $s_{T_{13}}$ to the logic nodes and computes $s_{T_{11}} = s_1 \oplus s_2$, $s_{T_{12}} = s_3 \oplus s_4 \oplus s_5$, and $s_{T_{13}} = s_6 \oplus s_7 \oplus s_8$. Next, $SN_1$ executes the same process to compute $K_{SN_1}$ as the subgroup key. Similarly, $SN_2$ and $SN_3$ compute $K_{SN_2}$ and $K_{SN_3}$, and BS computes $K_{BS}$ as GK. Each member stores all the keys along the path from the corresponding leaf to the root in the tree. $SN_1$ multicasts the following messages: $E_{k_{T_{11}}}(k_{SN_1})$, $E_{k_{T_{12}}}(k_{SN_1})$, $E_{k_{T_{13}}}(k_{SN_1})$ and $E_{k_{SN_1}}(k_{BS})$.

## 3.2 Group Membership Events

A comprehensive group key management scheme must provide key adjustment protocols stemming from membership changes. We distinguish among single and multiple member operations. Our protocol can handle the following four operations:

(1) Joining: a new member is added to the group.

(2) Leaving: a member is removed from the group.

(3) Mergence: multiple users are added to the group.

(4) Partition: multiple users are removed from the group.

To achieve both the backward secrecy and the forward secrecy, a new GK must be regenerated when membership changes. MDS code is used to achieve the dynamic key distribution. In order to weigh the communication cost as the number of encrypted messages to update the keys by SN, the following definition is borrowed [27]. The weight $w_i$ of a node $i$ is the sum of the degrees of all nodes on the path from node $i$ to the root. Supposing that the ancestor weight of the root BS is $w_r = 0$, $p$ is defined to be the parent of $i$ for $i \neq r$, and $\deg(p)$ is defined to be the degree for the node $p$. Then $w_r = w_p + \deg(p)$ is set. The tree weight $W(T)$ is the maximum degree in the tree such as $W(U_1) = W(U_2) = 0$, $W(T_{11}) = 2$, and $W(T_{12}) = W(T_{13}) = 3$. A node is weight-balanced if the node weight of its children differs by at most 1. A weight-balanced 2-3 tree is defined, where all nodes are weight-balanced and all internal nodes have degree 2 or 3. The following sections present the rekeying process and the tree adjustment rules when membership changes.

### 3.2.1 A Sensor Node's Joining

When a sensor node joins the network, GK is renewed to satisfy the backward secrecy. In order to reduce the communication cost and keep the tree balanced, the joing sensor node should be inserted into the node whose children are leaves with the minimum ancestor weight $W(T)$ in the tree. For example, the rekeying process is followed when a new user $u_{17}$ joins the group in Fig. 3.

(1) The minimum weight of the whole tree is $W(T_{11}) = 2$, therefore $u_{17}$ will be inserted into the subtree whose root is $T_{11}$.

(2) $SN_1$ assigns a seed key for $u_{17}$.

(3) $SN_1$ regenerates the new session key $K_{T_{11}}'$ for $u_1, u_2$ and $u_{17}$.

(4) $BS$ regenerates the new $GK'$.

(5) The following messages $E_{k_{BS}}(k_{BS}')$, $E_{k_{T_{11}}'}(k_{SN_1}')$ and $E_{k_{T_{11}}'}(k_{BS}')$ are multicasted.

Because the joing sensor node should be inserted into the node whose children are leaves with the minimum ancestor weight $W(T)$ in the tree, the node weight of its children differs by at most 1. We show rules for insertion in Fig. 4. We borrow the following definitions on the tree properties [27, 29]. For each rule, the left part shows the subtree before applying the rule and the right part shows the subtree after applying the rule. The number above the arrow shows the change in the node weight of the subtree root and the number below the arrow shows the communication cost for applying the rule. The upper right of every node shows the node weight. The symbol "/" means "or", for example 4/5 means the weight may be 4 or 5. Note that some of the nodes do not show the node weight, which means that the node weight is not constrained and can be chosen to be any value that maintains weight-balanced with its siblings. In Fig. 4(d), the black node represents a pseudo-node used to reduce the communication cost and keep the tree weight-balanced when membership changes. In the other rules (a), (b), (c), and (e), pseudo-nodes are not necessary because the tree can keep balanced with the existing sensor nodes. Because insertions are always done at a point that causes the minimal increase in tree weight, the tree is still weight-balanced.

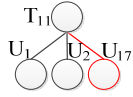

Fig. 3. A node's joining

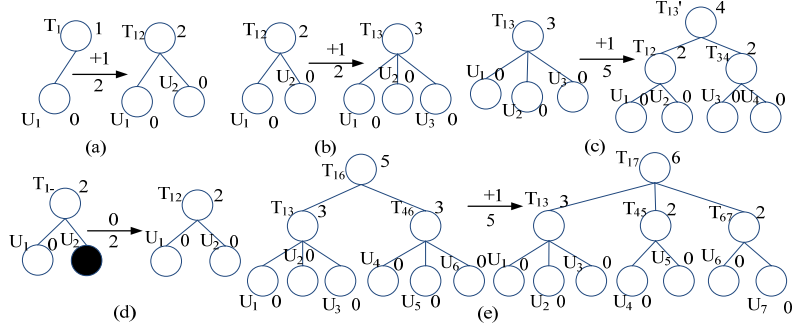

Fig. 4. The adjustment rules corresponding to a node's joining

### 3.2.2 A Sensor Node's Leaving

GK must be renewed to satisfy the forward secrecy when a sensor node leaves. The position cases where the removed node may locate are shown in Fig. 5. The deletion of a single node will cause imbalance in the tree. In our protocol, the virtual nodes, i.e. pseudo-nodes will be filled into the positions of the tree leaves to keep the tree weight-balanced and minimize the key update operations. The adjustment rules are summarized in Fig. 6 and Fig. 7. The adjustment rules for the bottom layer led by the imbalance are shown in Fig. 6, and the rules for the middle layer are shown in Fig. 7. These rules can be applied recursively.

For example $u_{18}$ will be removed in Fig. 8, the following steps illustrate how to apply the above rules and how to update GK. $u_{18}$'s position corresponds to the Fig. 5 (a) and a pseudo-node is added to the tree by the rule of Fig. 6 (a). $W(T_{11})$ is the same as before, so the adjustment for the middle layer is not necessary.

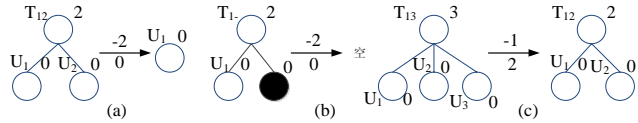

Fig. 5. The positions of a single node's leaving

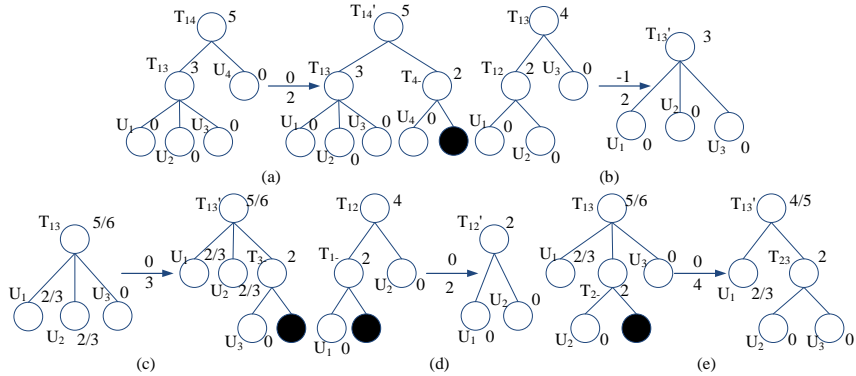

Fig. 6. The adjustment rules corresponding to a node's leaving in the bottom layer

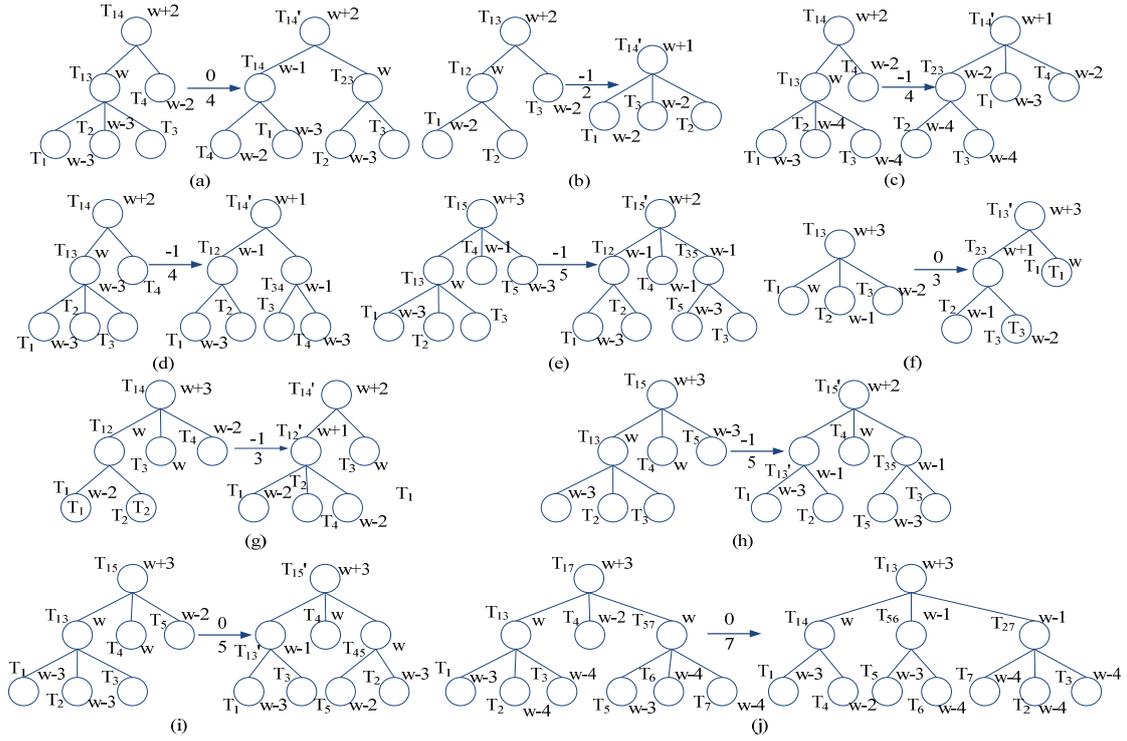

Fig. 7. The adjustment rules corresponding to a node's leaving in the middle layer

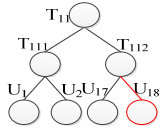

Fig. 8. An example of deleting $u_{18}$

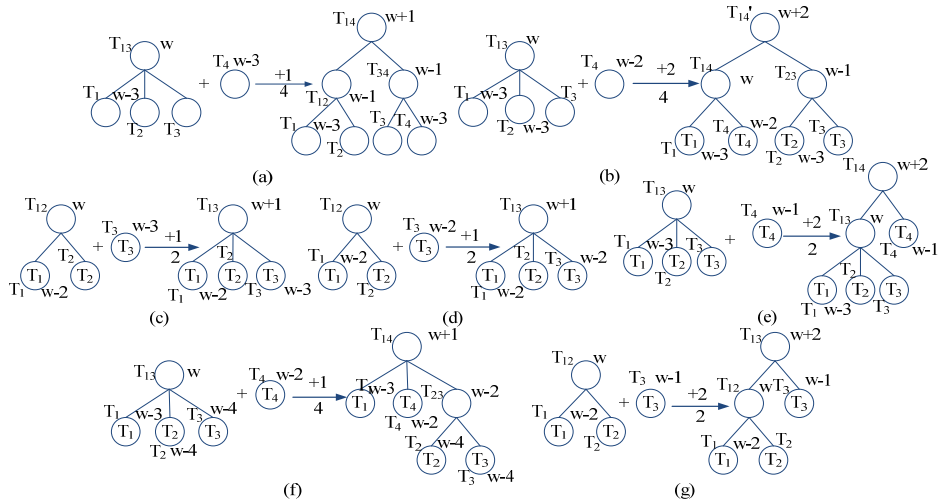

Fig. 9. The adjustment rules corresponding to mergence

### 3.2.3 Mergence

When multiple sensor nodes join the tree controlled by $SN_i$, $SN_i$ will construct these nodes into a weight-balanced 2-3 tree with the lowest tree weight. If the new tree weight is less than the old tree weight, the new tree will be merged into the old tree. On the contrary, the old tree will be merged into the new tree. By experiments, we can always find a node in the higher weight tree whose tree weight difference does not exceed three compared with the smaller weight tree. The adjustment rules are listed in Fig. 9. If the tree weight in the upper layers becomes imbalanced, the set of rules can be used in Fig. 7. The key update process has been mentioned in Section 3.2.

### 3.2.4 Partition

When multiple sensor nodes wants to leave the tree controlled by $SN_i$, $SN_i$ will delete these nodes one by one. The tree structure will be reconstructed after all nodes have left the group. The tree balance adjustments will be from bottom to top until all the subtrees are balanced. On the one hand, if the node tree weight of its children differs by less than 3, the adjustment rules will be classified into the bottom layer rules and the upper layer rules. In the bottom layer, some rules are shown in Fig. 6 and the other rules are shown in Fig. 10. In the upper layer, some rules are shown in Fig. 7 and the other rules are set in Fig. 11. On the other hand, if the node tree weight of its children differs by more than 3, the rules are the same as those in Fig. 9.

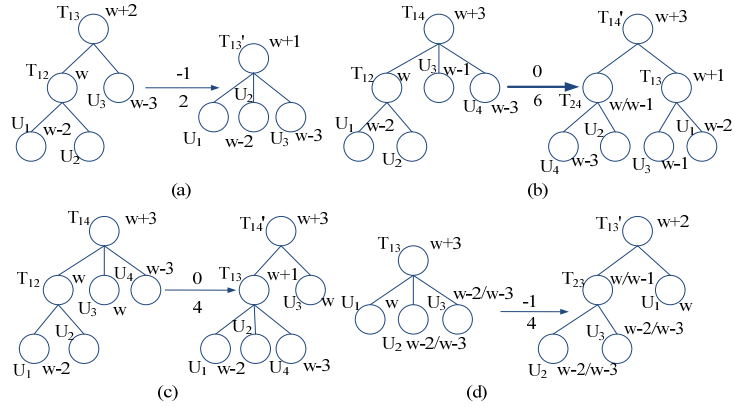

Fig. 10. The adjustment rules corresponding to partition in the upper layer

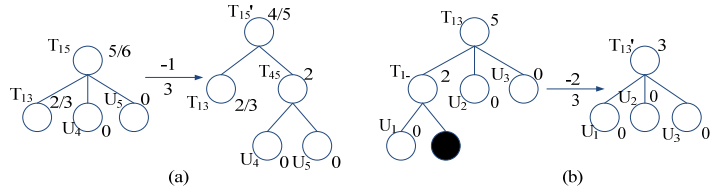

Fig. 11. The adjustment rules corresponding to partition in the lowest layer

## 4. Security and Performance Analysis

In this section, we give the security analysis in terms of guessing attack, key independence, forward secrecy, backward secrecy and conspiracy attack. Performance analysis are also given in terms of computation, storage and communication overhead. We compare our protocol with PCGR [12], GKD [10], and GKSS [22]. Symbols are used in Table 2.

### 4.1 Security Analysis

(1) *Guessing attack*: As mentioned above, *GK* is computed based on the random $r$ and seed keys of all members. Since $r$ and $m_2 m_3 ... m_n$ from $SN_i$ are multicasted in plaintext and thus known to all parties including unauthorized receivers who attempt to access the current session, the security of the new session key relies on the secrecy of a code word symbol $c_{j_i}$ that only authorized member $u_i$ can compute. Because the equation $GF(q): c_{j_i} = H(s_i \| r)$ is set, attackers can get GK only by the following three ways [25]: Brute-force attack, guessing a symbol $c_{j_i}$, and guessing $u_i$'s seed key.

Table 2. Notations of analysis

| Symbol | Meaning |
| --- | --- |
| $t$ | The order of polynomial |
| $L$ | the length of GK |
| $l_r$ | the length of a random number $r$ |
| $l_j$ | the length of $j_i$, the maximum length is $m$ |
| $h$ | Tree's height |
| E/D | Encryption/ Decryption operation |
| $n$ | Number of leaves |
| $n_1$ | The number of other nodes except the leaves, its value is within $[2^{h-1}-1, (3^{h-1}-1)/2]$ |
| $\deg(x)$ | The degree of the node $x$ |
| M/U | Multicast/ Unicast |
| $C_E, C_D, C_H, C_{EP}$ | The computation cost of one evaluation of encryption, decryption, hash function and, the module exponentiation. |
| $C_M$ | The computation cost of the matrix multiplication |

The effort that an attacker makes to deduce GK depends on the following parameters, namely, the size of finite field $GF(2^m)$, $t$, $l_r$ and $l$. The longer the size of these parameters is $l$, the more difficultly GK can be guessed. The exact sizes of these parameters can ensure the security of GK. If $m = t = l_r = l$ is satisfied, the effort that an attacker needs to deduce a GK is no less than that of a brute-force attack, which can be proved by the following three proofs.

① The entropy of GK is $H(GK) = \log_2 GK = l$.

② The entropy of $c_{j_i}$ is $H(c_{j_i}) = \log_2 2^m = m = l$. If an attacker wants to infer GK by guessing $c_{j_i}$, he must first pick up an arbitrary location $j_i$ where $n \leq j_i \leq n + n_1$.

③ The entropy of $s_i$ is $H(s_j) = \log_2 2^t = t = m = l$, while the entropy of $S_i$ including $s_i$ and $j_i$ is $H(S_i) = H(s_i) + H(j_i) = l + \log_2 L$. Considering the fact that $r$ is also known to the attacker, the effort that an attacker needs to deduce $s_i$ from $r$ and $c_{j_i}$ is the conditional entropy $H(s_i | r, c_{j_i}) = H(s_i) = l$.

Combining all the above possible attacks, the amount of information to guess a true $l$-bit GK is at least $l$ bits, i.e., the attacker needs to make as much effort as a brute-force attack to deduce the right GK.

(2) *Confidentiality of communication*: Multicast messages are encrypted by GK known by legal sensor nodes.

(3) *Forward security*: A leaving member can not access the subquent messages after before it leaves the group.

. (4) *Backward security*: A new member can not decode messages exchanged before it joins the group.

. (5) *Group Key Secrecy*: In Section 3, it has expounded that only legal members can gain GK, therefore it is computationally infeasible for a passive adversary to discover GK.

(6) *Key Independence*: The freshness of GK is decided by the random number $r$, thus GK is independent because of the independent $r$. The number of GK is decided by $l_r$, i.e. $2^{l_r}$.

(7) *Conspiracy attack:* A conspiracy attack means that some old members cooperate to deduce the current GK. Since the independence, the forward secrecy and the backward secrecy of GK can be guaranteed, one possible way for an old member to get a new GK is to calculate the seed key of a current member from old keys. It is easy to compute a symbol $c_{j_i}$ based on the equation (1). $c_{j_i} = H(s_i \| r)$ is set, therefore it is impossible or at least computationally hard to compute a seed key $s_i$ even if $c_{j_i}$ is known.

## 4.2 Performance Analysis

In this section, performance analysis are focus on the computation, communication and storage overhead of every sink node and every sensor node. We analyze from the group initialization and the four operations of joing, leaving, mergence and partition.

### 4.2.1 Performance Analysis on Group Initialization

(1) Communication complexity

When a new member $u_i$ is authorized to join a multicast group, its $SN$ will assign a seed key pair $(j_i, s_i)$ to it securely. The number of unicast messages comes to $nU$.

$SN$ multicasts $r$ and $m_2 m_3 ... m_n$ to every member in plaintext whose length is $l_r + (n-1)m$. As already discussed, it is secure enough to set $m = t = l_r = l$. Thus, the length of multicast message is $l_r + (n-1)m = nl$ if the length of $m_i$ is $m$ bit.

$SN$ multicasts the keys in the key path encrypted by the node keys in the penultimate layer. The total number in the penultimate layer is $[2^{h-2}, 3^{h-2}]$, therefore the total number of multicast messages is $[2^{h-2}, 3^{h-2}] \times 2M$.

(2) Computation complexity

$SN$ unicasts a seed key pair $(j_i, s_i)$ to every member securely, hence the encryption operations of $SN$ are $nC_E$. $SN$ computes $GF(q): c_{j_i} = H(s_i \| r)$ for every node, hence the hash operations are $[2^h - 1, (3^h - 1)/2]C_H$, i.e., $(n_1 + n)C_H$.

Every node performs matrix operations to recover GK or the subgroup key, hence the computation cost is $n_1 C_M$. Every sensor node decrypts to get its seed key and its keys in the key path, i.e., $C_H + 2C_D$.

(3) Storage complexity

$SN$ needs to store the seed keys of all members, i.e., $n(\lceil \log_2 L \rceil + t)$. Since it is secure to set $m = t = l_r = l$, and the maximum of $L$ is $2^m + 1$ [28], $n(\lceil \log_2 L \rceil + t) = n(2l + 1)$ can be inferred. Moreover, $SN$ needs $n_1 l_r$ bits to store the keys.

A current member $u_i$ only needs $(2l+1)$ bits to store its seed key and $(h-1)l_r$ bits to store keys in its key path.

### 4.2.2 Performance Analysis on A Node's Joining

(1) Communication complexity

$SN$ unicasts a seed key pair to the new member securely. $SN$ multicasts once in order to let the new member compute the new sub-group key. The keys from the insertion node to the root should be updated, therefore $SN$ needs $(h-2)M$. $SN$ multicast the new GK to all the old members. In short, the total communication cost is $hM+U$.

(2) Computation complexity

$SN$ unicasts a seed key pair $(j_i, s_i)$ to every new member securely, therefore the computation cost is $C_E$. The keys from the insertion node to the root should be updated, hence $SN$ needs $(h-1)C_M$ and $[2(h-1), 3(h-1)]C_H$.

Every new member needs one $C_H$ to get $c_{j_i}$, one $C_D$ to get its seed key and one $C_D$ to get its keys in the key path, i.e, $C_H + 2C_D$. Every old member needs one $C_D$ to get the new GK.

(3) Storage complexity

$SN$ needs $n(\lceil \log_2 L \rceil + t) = n(2l+1)$ bits to store seed keys for members and $n_1 l_r$ bits to store the keys of the nodes.

A current member $u_i$ only needs $(2l+1)$ bits to store its seed key and $(h-1)l_r$ bits to store keys in its key path.

### 4.2.3 Performance Analysis on A Node's Leaving

(1) Communication complexity

The tree should be reconstructed. $SN$ needs to $[2^{h-2}, 3^{h-2}]M + [\deg(S_i)-1]M$ to send the new GK with $h \geq 3$.

(2) Computation complexity

The keys from the deletion node to the root should be updated, hence $SN$ needs $(h-1)C_M$ and $[2(h-1), 3(h-1)]C_H$. Some nodes whose positions change need $C_H + C_D$, while other members need one $C_D$ to get the new GK.

(3) Storage complexity

$SN$ needs $n(2l+1)$ bits to store seed keys for members and $n_1 l_r$ bits to store the keys of the nodes.

A current member $u_i$ only needs $(2l+1)$ bits to store its seed key and $(h-1)l_r$ bits to store keys in its key path.

### 4.2.4 Performance Analysis on Mergence

When multiple sensor nodes join a tree, these multiple sensor nodes will be reconstructed a weight-balanced tree. In order to facilitate discussion, supposing the height of the new tree is $h_1$, the height of the old tree is $h_2$ and $h_1 \leq h_2$ is set, the new tree will be merged into the old tree. After mergence, the tree height is $h_2 + 1$ with $n$ nodes. If the tree with the height $h_1$ is seen as a logic node, the most complex case is that the positions of all the nodes in the tree branches of the logic node's grandfather $p$ will change.

(1) Communication complexity

The communication complexity of the tree initialization with the height $h_1$ is $xU + M + [2^{h_1-2}, 3^{h_1-2}]M$.

For the tree with height $h_2$, the new GK encrypted by the old GK is multicasted. The tree with height $h_1$ is seen as a logic node, and the new GK of its sibling brothers (2/3) encrypted by the old GK is multicasted. At the same time, the keys from its $p$ to the root need update. The total communication complexity is $(h - h_1)M + (1 + 2/3)M$, i.e., $(h - h_1 + 3/4)M$.

(2) Computation complexity

Supposing that there are $x$ members in the tree of $h_1$, $SN$ needs $[2^{h_1+2}, 3^{h_1+2}]C_H$ to merge these two trees into one. $SN$ needs $xC_E$ to send the seed key pairs securely. $SN$ needs $[2^{h_1-2}, 3^{h_1-2}]M + (h - h_1)M$ to multicast the keys of the logic nodes.

Every new member needs one $C_H$ to get $c_{j_i}$, one $C_D$ to get its seed key and one $C_D$ to get its keys in the key path, i.e, $C_H + 2C_D$. Every old needs one $C_D$ to get the new GK.

(3) Storage complexity

$SN$ needs $n(2l+1)$ bits to store seed keys for members and $n_1 l_r$ bits to store the keys of the nodes.

A current member $u_i$ only needs $(2l+1)$ bits to store its seed key and $(h-1)l_r$ bits to store keys in its key path.

### 4.2.5 Performance Analysis on Partition

The most complicated case is the same as the group re-initialization due to the irregularities of partition. In this case, the performance anylysis are the same as those of the initialization.

### 4.2.6 Performance Comparison

In this section, our protocol is compared with PCGR [12] and GKD [10] in terms of storage, computation and communication overhead. Our protocol belongs to the decentralized management protocol, while PCGR and GKD belong to the distributed key management protocol. The comparison results are shown from Table 3 to Table 5. From these tables, we can see that the performance of our protocol is superior to PCGR and GKD. With the increasement of leaf nodes, the storage overhead in our protocol is proportional to the tree height, while computation and communication cost is constant as shown in Fig. 12. The storage, computation and communication of PCGR and GKD are proportional to the number of tree leaves as shown in Fig. 12.

Table 3. The comparison on the group initialization among our protocol, PCGR and GKD

|  | Our protocol | PCGR | GKD |
| --- | --- | --- | --- |
| Storage overhead | $(2L+1) + (h-1)L$ | $(n+1)(t+1)L$ | $(2t+3+w)L$ |
| Computation overhead | $C_H + 2C_D$ | $O((n+1)t^2) + nC_E$ | $O(2t^2)$ |
| Communication overhead | 0 | $n(t+1)L$ | $(\frac{5}{2}t+1)n_B L$ |

Table 4. The comparison on a node's joining among our protocol, PCGR and GKD

|  | Our protocol | PCGR | GKD |
|---|---|---|---|
| Storage overhead | $(2L+1)+(h-1)L$ | $(n+1)(t+1)L$ | $(m+1)L$ |
| Computation overhead | $C_H + 2C_D$ | $O(\mu^3) + nC_E$ | $O(t^3) + nC_E$ |
| Communication overhead | 0 | $nL$ | $(t+1)L$ |

Table 5. The comparison on a node's leaving among our protocol, PCGR and GKD

|  | Our protocol | B-PCGR | GKD |
|---|---|---|---|
| Storage overhead | $(2L+1)+(h-1)L$ | $(n+1)(t+1)L$ | $(m+1)L$ |
| Computation overhead | $C_H + C_D$ | $O(\mu^3) + nC_E$ | $O(t^3) + nC_E$ |
| Communication overhead | 0 | $nL$ | $(t+1)L$ |

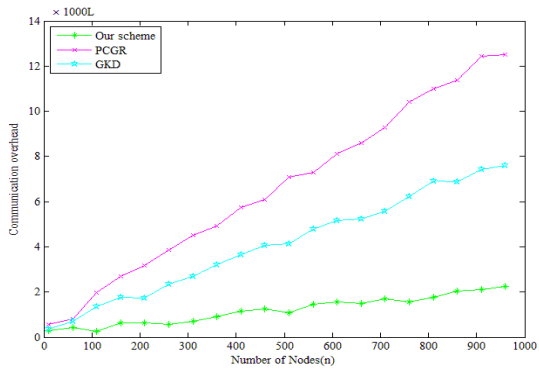 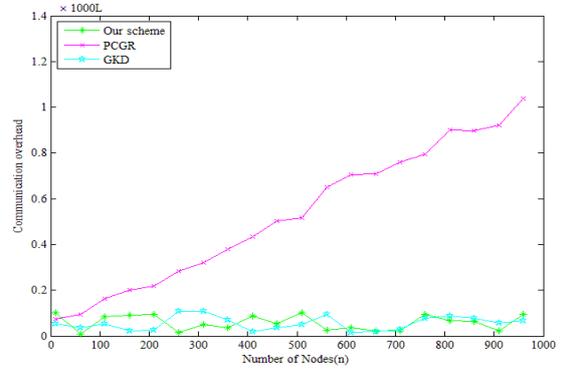

(a) The comparison on the group initialization    (b) The comparison on group rekeying

Fig. 12. The comparison on the group initialization

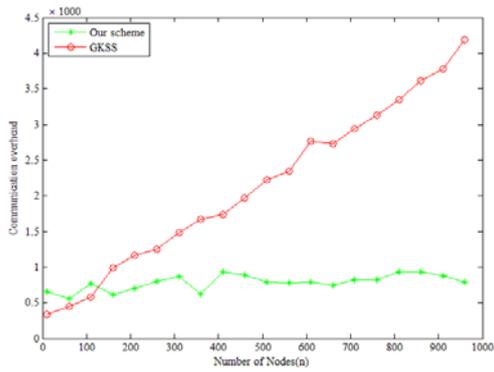 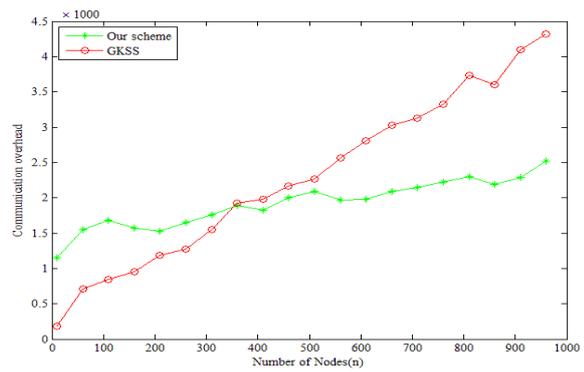

(a) The comparison of adding a member    (b) The comparison of deleting a member

Fig. 13. The comparison of adding a member

GKSS [22] belongs to the distributed key management protocol as our protocol. The secret shares are calculated by the cluster heads which choose the one-order polynomial to improve efficiency. Our scheme is compared with GKSS Table 6 to Table 8. From these tables, we can see that the storage of the sink node and the ordinary node in our protocol and GKSS is almost identical, which varies directly with the number of tree leaves as shown in Fig.

13. When GK is updated, the computation and communication of our protocol is proportional to the tree height and the computation and communication of GKSS is proportional to the number of leaves, which shows that our protocol is superior to GKSS as shown in Fig. 13.

Table 6. The comparison on initializing the group between our protocol and GKSS

|  | Our protocol | | GKSS | |
| --- | --- | --- | --- | --- |
|  | SN | The ordinary node | SN | The ordinary node |
| Storage overhead | $n(2L+1)+n_1 L$ | $(2L+1)+(h-1)L$ | $(2n+3)L$ | $4L$ |
| Computation overhead | $nC_E + (n_1+n)C_H + n_1 C_M$ | $C_H + 2C_D$ | $nC_E$ | $C_D$ |
| Communication overhead | $2nL$ | 0 | $(n+m)L$ | 0 |

Table 7. The comparison on a node's joining between our protocol and GKSS

|  | Our protocol | | GKSS | |
| --- | --- | --- | --- | --- |
|  | SN | The ordinary node | SN | The ordinary node |
| Storage overhead | $n(2L+1)+n_1 L$ | $(2L+1)+(h-1)L$ | $(2n+3)L$ | $4L$ |
| Computation overhead | $C_E + [2(h-1), 3(h-1)]C_H + (h-1)C_M$ | $C_H + 2C_D$ | $O(2^2) + nC_E$ | $O(1) + C_D$ |
| Communication overhead | $[h+3, h+4]L$ | 0 | $4(n+h)n$ | 0 |

Table 8. The comparison on a node's leaving between our protocol and GKSS

|  | Our protocol | | GKSS | |
| --- | --- | --- | --- | --- |
|  | SN | The ordinary node | SN | The ordinary node |
| Storage overhead | $n(2L+1)+n_1 L$ | $(2L+1)+(h-1)L$ | $(2n+3)L$ | $4L$ |
| Computation overhead | $[2(h-1), 3(h-1)]C_H + (h-1)C_M$ | $C_H + C_D$ | $O(2^2) + nC_E$ | $O(1) + C_D$ |
| Communication overhead | $[2^{h-2}, 3^{h-2}]L + [\deg(S_i)-1]L + L$ | 0 | $2L + 4(n+h)n$ | 0 |

## 5. Conclusion

In this paper, we present a group multicast key management protocol based on the weight-balanced 2-3 tree. In our protocol, the whole group is divided into two layers. The decentralized group key management is adopted, while the whole group is organized as an irregular tree and every subgroup is organized as a weight-balanced 2-3 tree. Hence, our scheme takes advantages of both centralized group key management method and distributed key group management method. Compared with some other protocols, our protocol shows higher superiority on security, scalability and performance, which is more suitable for WSNs.

As part of our future work, we intend to do more research on fault-tolerant features for scheme in order to prevent some of communication messages from being lost during the key distribution and updating. We will also evaluate our scheme by simulations and test-bed experiments.

## 6. Acknowledgements

This work was partially supported by the Natural Science Foundation of China under Grants No.60703101 and

**References**

[1] Akyildiz, I. F., Su, W., Sankarasubramaniam, Y. and Cayirci, E., Wireless sensor networks: a survey. *Journal of Computer Networks*, 38(2002), 393-422.

[2] Gaddour, O., Koubaa, A. and Abid, M., SeGCom: A secure group communication mechanism in cluster-tree wireless sensor networks. *The First International Conference on Networks & Communications*, Chennai, 2009, pp. 1-7.

[3] Klaoudatou, E., Konstantinou, E., Kambourakis, G. and Gritzalis, S., A Survey on Cluster-Based Group Key Agreement Protocols for WSNs. *IEEE Communications Surveys & Tutorials*, PP (2010), 1-14.

[4] Pietro, R. D., Michiardi, P. and Molva, R., Confidentiality and integrity for data aggregation in wsn using peer monitoring. *Security and Communication Networks*, 2(2009), 181-194.

[5] Rafaeli, S., Hutchison, D., A Survey of Key Management for Secure Group Communication. *ACM Computing Surveys*, 35(2003), 309-329.

[6] Judge. P. and Ammar, M., Security Issues and Solutions in Multicast Content Distribution: A Survey. *IEEE Network*, 17 (2003), 30-36.

[7] Sun, H. M., Chen, C. M. and Chu, F. Y., An Efficient and Scalable Key Management Protocol for Secure Group Communications in Wireless Sensor Networks. *12th IEEE Symposium on Computers and communications*, Aveiro, 2007, pp. 495-500.

[8] Wong, C. K., Gouda, M. and Lam, S. S., Secure group communications using key graphs. *IEEE/ACM Trans. Networking*, 8(2000), 16-30.

[9] Di Pietro, R., Mancini, L. V., Law, Y. W., Etalle, S. and Havinga, P., LKHW: a directed diffusion-based secure multicast scheme for wireless sensor networks. *2003 International Conference on Parallel Processing Workshops (ICPPW'03)*, IEEE, Kaohsiung, 2003, pp. 397-406.

[10] Chandha, A., Liu Y. and Das, S. K., Group key distribution via local collaboration in wireless sensor networks. *Proc. 2nd Annual IEEE Communications Society Conf. Sensor and Ad HocCommunications and Networks*, California, 2005, pp.46-54.

[11] Panja, B., Madria, S. K. and Bhargava, B. K., Energy and communication efficient group key management protocol for hierarchical sensor networks. *IEEE International Conference on Sensor Networks, Ubiquitous and Trustworthy Computing*, Taichung, 2006, pp.384-393.

[12] Zhang, W. S. and Cao, G. H., Group rekeying for filtering false data in sensor networks: a predistribution and local collaboration based approach. *In Infocom IEEE*, 2005, pp. 503-514.

[13] Eltoweissy, M., Younis, M. and Ghumman, K., Lightweight key management for wireless sensor networks. *2004 IEEE International Conference on Performance, Computing, and Communications*, 2004, pp. 813-818.

[14] Wadaa, A., Olariu, S. and Wilson, L., Scalable cryptographic key management in wireless sensor networks. *24th Int. Conf: Distributed Computing Systems Workshops*, Washington, 2004, pp. 796-802.

[15] Morales, L., Sudborough, I. H., Eltoweissy, M. and Heydari, M. H., Combinatorial optimization of multicast key management. *36th Annual Hawaii Int. Conf: System Sciences*, Washington, 2003, pp. 33-50.

[16] Redwine, S. T., Jr., A logic for the exclusion basis system. *37th Annual Hawaii International Conf: on System Sciences*, IEEE Computer Society, Washington, 2004, pp.280-285.

[17] Thepvilojanapong, N., Tobe, Y. and Sezaki, K., A proposal of secure group communication for wireless sensor networks. *The 23th Computer Security (CSEC) Group Meeting, IPSJ*, Tokyo, 2003, pp. 47-52.

[18] Tubaishat, M., Yin, J., Panja, B. and Madria, S., A secure hierarchical model for sensor network. *SIGMOD Rec.*, 33(2004), 7-13.

[19] Wang, Y. and Ramamurthy, B., Group Rekeying Schemes for Secure Group Communication in Wireless Sensor Networks. *IEEE International conference on communications*, 2007, pp. 3419-3424.

[20] Zhang L. P., Wang Y. and Li G. L., A Novel Group Key Agreement Protocol for Wireless Sensor Networks. *International Conference on Wireless Communications & Signal Processing*, Nanjing, 2009, pp. 1-4.

[21] Li, G., Wang, Y. and He, J. S., Efficient Group Key Management Scheme in Wireless Sensor Networks. *IITSI2010*, Jinggangshan, 2010, pp.411-415.

[22] Yang, X. Y., Wang, L. Q. and Zhang, Q., A New Group Key Management Protocol in WSNs Based on Secret Sharing. *Power and Energy Engineering Conference,* Chengdu, 2010, pp. 1-4.

[23] Poornima, A. S. and Amberker, B. B., A Secure Group Key Management Scheme for Sensor Networks. *Fifth International Conference on Information Technology: New Generations*, Las Vegas, 2008, pp. 744-748.

[24] Raj, S. B. E. and Lalith, J. J., A Novel Approach for Computation-Efficient Rekeying for Multicast Key Distribution. *International Journal of Computer Science and Network Security*, 2009, pp. 279-284.

[25] Xu, L. H. and Huang, C., Computation Efficient Multicast Key Distribution. *IEEE Trans. Parallel Distributed Systems*, 19(2008), 577-587.

[26] Plank J. S. and Ding Y., Correction to the 1997 Tutorial on Reed Solomon Coding. *Software, Practice and Experience*, 35(2005), 189-194.

[27] Snoeyink, J., Suri, S. and Varghese, G., A Lower Bound for Multicast Key Distribution. *IEEE INFOCOM*, Alaska, 2001, pp. 1-10.

[28] Goshi J. and Ladner, R. E, Algorithm for Dynamic Multicast Key Distribution. *ACM Journal of Experimental Algorithmics*, ACM, Zurich, 2006, pp. 1-37.

[29] Goshi, J. and Ladner, R. E., Algorithms for dynamic multicast key distribution trees. *Proceedings of the twenty-second annual symposium on Principles of distributed computing*, ACM, 2003, pp. 243-251.

[30] Yao, L., Wang, L., Kong, X. W., Wu, G. W. and Xia F., An inter-domain authentication scheme for pervasive computing environment. *Computers and Mathematics with Applications*, 60(2010), 234-244.

*Corresponding author: Feng Xia, Ph.D.

School of Software, Dalian University of Technology, Dalian 116620, China

E-mail: f.xia@ieee.org